\documentclass[12pt]{iopart}
\usepackage{graphicx}
\usepackage{hyperref}
\usepackage{epsfig}
\usepackage{multicol}
\usepackage{color}
\usepackage{dcolumn}
\usepackage{bm}
\usepackage{epstopdf}

\begin{document}
\title{Efimov three-body states on top of a Fermi sea}
\author{Nicolai Gayle Nygaard and Nikolaj Thomas Zinner}

\address{Lundbeck Foundation Theoretical Center for Quantum System Research,
Department of Physics and Astronomy, Aarhus University, DK-8000 Aarhus C, Denmark}
\date{\today}

\begin{abstract} 
The stabilization of Cooper pairs of bound electrons in the background of a Fermi sea is the origin
of superconductivity and the paradigmatic example of the striking influence of many-body physics 
on few-body properties. In the quantum-mechanical three-body problem the famous Efimov effect yields
unexpected scaling relations among a tower of universal states. These seemingly unrelated problems
can now be studied in the same setup thanks to the success of ultracold atomic gas
experiments.
In light of the tremendous effect of a background
Fermi sea on two-body properties, a natural question is whether a background can modify or even destroy
the Efimov effect.
Here we demonstrate how the generic problem 
of three interacting particles changes when one particle is embedded in 
a background Fermi sea, and show that Efimov scaling 
persists. It is found in a scaling that relates 
the three-body physics to the background density of fermionic
particles.
\end{abstract}

\maketitle

The discovery by Vitaly Efimov of an anomaly in the spectrum of 
three equal mass particles when a two-body bound state is at zero energy \cite{efi70} 
has excited theorists and experimentalist alike in the past four decades. 
While much effort was put into detecting these universal bound states
within nuclear and molecular physics, the search has been unsuccessful
so far \cite{jen04}. Only with the emergence of ultracold atomic gases
and experimental control of the two-body interaction \cite{bloch2008}, was the 
prediction of Efimov finally observed in cold gases of alkali 
atoms \cite{kraemer2006,ferlaino2009,barontini2009,pollack2009,zaccanti2009,gross2009,huckans2009,lompe2010,nakajima2011,machtey2012,rem2013,roy2013}.
The observations showed that one indeed finds a universal 
geometric scaling law for the three-body states. This 
law states that an infinite tower of three-body bound states exist 
and that the energy is smaller by a factor of $22.7^2\sim 515$, as one moves from one state
to the next, approaching zero energy. This implies that the observation of 
such states is extremely difficult, highlighting the enormous achievement 
of the ultracold gas experiments of recent years.

From the original idea of Cooper pairs \cite{cooper1956}, that are an essential ingredient in 
the famous Bardeen-Cooper-Schrieffer theory of superconductivity \cite{bcs1957}, it
is clear that few-body properties can be strongly influenced by a many-body background.
In that case a bound pair of opposite spin electrons can form in the presence of a Fermi sea, even if the pair would be unstable 
without the many-body background. A natural question that appears
in this context is whether this binding and the 
macroscopic coherence
of the pairs in general can survive
if we start to polarize the system, i.e. change the size of the Fermi sea for 
one of the electronic spin states but not the other. This gives rise to the study of exotic
pairing phases and in the extreme limit of just a single impurity in a Fermi sea polaronic physics emerges. In ultracold atomic gases, the spin components are mimicked by internal hyperfine states whose relative population can be adjusted freely, and experiments have provided 
numerous insights into the interplay between the two-body problem and many-body physics \cite{hulet,ketterle,zwierlein,zwierlein2,salomon}.

\begin{figure*}[ht!]
\centering
\includegraphics[scale=0.7]{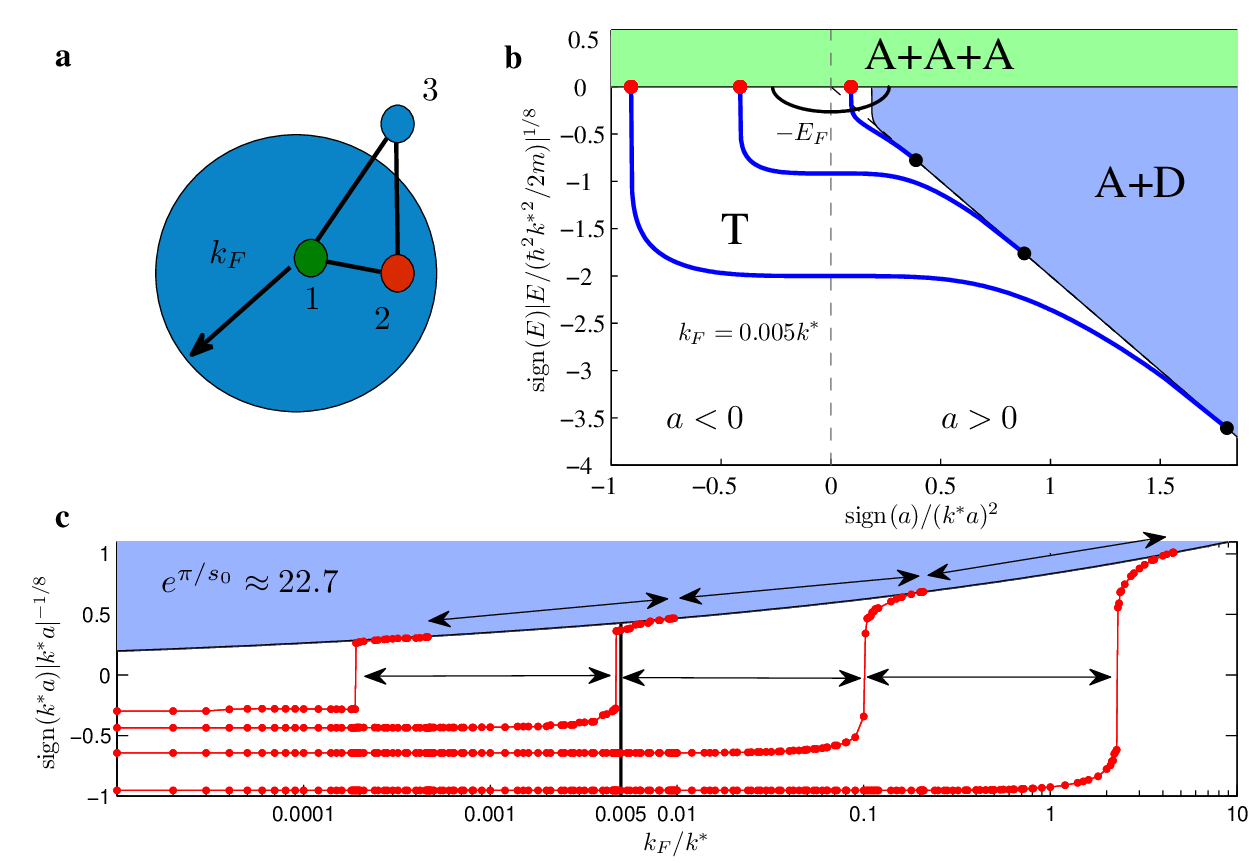}
\caption{Efimov spectrum in the presence of a Fermi sea. {\bf a} Schematic picture
of the three interacting particles with a Fermi sea in the third component (blue filled circle). {\bf b} 
Solid blue lines
show the binding energies of the Efimov trimers (T), while the thin black lines mark the beginning of the A+A+A three-atom 
(upper) and A+D atom-dimer (right) continua. The intersection of the Efimov states with these continua are marked with red and black dots, respectively. The Fermi wave number is $k_{\rm{F}}=0.005k^*$. Here $k^*$ is the three-body 
parameter. Notice in particular how the 
least bound Efimov trimer has moved to the right of the resonance where the scattering length diverges and 
is a well-defined three-body bound state in spite of the presence of a bound two-body dimer. A thick black 
line marks the region where the Fermi energy is a relevant energy scale and influences the spectrum. This region grows with $k_{\rm{F}}$ and modifies the trimers to move towards the 
atom-dimer threshold roughly when they are inside the circle. Dashed lines mark the resonance (vertical) and 
the dimer threshold in vacuum  (diagonal). {\bf c} The spectrum 
flow of the Efimov trimers with zero three-body energy as function of $k_{\rm{F}}$. The solid black line indicates the position of the two-body atom-dimer threshold. The 
vertical axis has been scaled for visibility. The successive trimer thresholds scale with $e^{\pi/s_0}\sim 22.7$ 
for small $k_{\rm{F}}/k^*$, when they cross the resonance position as $k_{\rm{F}}$ increases, and again as they merge with 
the atom-dimer continuum.\label{spec}
}
\end{figure*}

Here we address the question of what happens when one merges three-body physics
and the Efimov effect with a many-body background. We consider
a system of three equal mass, but distinct, particles with one of the particles being immersed in 
a Fermi sea background as illustrated in figure~\ref{spec}a.
The latter means that the Pauli exclusion principle must be 
taken into account when calculating the three-body properties. 
The system we consider
can in fact be related to the polaron problem and corresponds to having two 
impurities that interact with a Fermi sea.
While we have chosen 
the simplest case of a single Fermi sea in this initial study, this is not an essential 
assumption. Our method can be applied also to the case with two or three Fermi seas.

As we will now discuss, the universal scaling laws, originally obtained 
by Efimov, hold also in the presence of a Fermi sea. Furthermore, we find a new scaling 
law related to the background Fermi energy, which may be interpreted as a
many-body scaling for the three-body bound states. Our results demonstrate that the Efimov effect
is robust in a many-body background. We consider first the generic
problem of Efimov; three particles interacting with the same two-body
interaction which is characterized by the scattering length, $a$, in the universal regime 
($|a|\gg r_0$ with $r_0$ the characteristic potential range). Afterwards, we discuss the three-body 
bound states in the three-component $^{6}$Li system that has been recently realized
\cite{huckans2009,lompe2010,nakajima2011}. For the latter system, we expect our 
findings to be accessible at or slightly above the densities of current experiments.

Our approach to the problem of three-body states in the presence of a background 
Fermi sea uses the momentum-space equations originally formulated by
Skorniakov and ter-Martirosian (STM) \cite{STM,danilov,pri}.
In the absence of a Fermi sea, it has been shown \cite{braham2006,pri} that our approach captures the full 
physics of the Efimov effect, and in vacuum it is equivalent to the 
Faddeev decomposition in the hyperspherical formalism applied to the 
lowest hyperradial potential which also captures the Efimov effect \cite{jen04}.
We consider the case where one of the three particles forming the bound state has a Fermi sea with
Fermi wave number, $k_{\rm{F}}$, in the limit of an inert sea with
no particle-hole pairs. We will address the consequences of this approximation 
and also the case of multiple Fermi seas below. Since we are interested in the 
universal properties of the three-body spectrum we work exclusively with 
zero-range interaction that can be parametrized by $a$,
but note that finite range corrections are easily 
taken into account in the STM approach. The scale of the spectrum is set by the three-body parameter $k^*$ \cite{braham2006}. 
Other recent studies considered similar issues in the 
Born-Oppenheimer limit \cite{nishida2009,macneill2011}, where the fermion mass is much smaller than the 
two impurities, and using variational wave functions \cite{mathy2011} inspired
by the original work on the two-body Fermi polaron problem by Chevy \cite{chevy}. 
The spectrum for $1/a=0$ as function of $k_{\rm{F}}$ was obtained 
using a semi-classical approximation in the Born-Oppenheimer limit \cite{macneill2011}.
Our approach using the STM equation is, however,
general and works for any choice of masses and any value of the 
interaction strengths. An independent study of Pauli blocking effects 
was presented in Ref.~\cite{niemann2012}. That study did not address the 
scaling relations of Efimov states in the presence of a Fermi sea.

In the presence of a Fermi sea, the 
two-body dimer threshold moves from $1/a=0$ to the $a>0$ side
of the two-body scattering resonance \cite{schuck} as seen in figure~\ref{spec}b. Based on 
this observation, it is natural to speculate that the three-body bound state
threshold with the continuum will also be moved as function of $k_{\rm{F}}$. 
We indeed confirm this 
expectation, but find that the states move in a fashion that respects
the original scaling found by Efimov. Note that the the Efimov 
scaling is indeed violated for non-zero $k_{\rm{F}}$. This is clearly 
seen in figure~\ref{spec}b where the first and second Efimov trimers 
have 22.7 scaling, while the third one does not (see also figure~\ref{spec}c).
In figure~\ref{spec}b we have
chosen a value of $k_{\rm{F}}$ such that a bound three-body state merges 
with the three-body continuum at zero energy for $a>0$. This is perhaps puzzling
since in the case of a single Fermi sea, a dimer could be formed between 
the two particles that do not have a background. However, we find numerically
that the three-body state is well-defined in this region, i.e.
the three-body energy has a vanishingly small imaginary part, and Pauli
blocking thus seems to stabilize the system from decay into
a dimer and an atom.

The famous Efimov scaling factor, $e^{\pi/s_0}\sim 22.7$, where $s_0=1.00624$ for 
equal mass particles,
can be seen for $k_{\rm{F}}=0$ both on resonance ($1/a=0$) and at the thresholds on either side as a 
factor determining the ratio of energies and (dis)-appearance points \cite{braham2006,nie01}.
However, when we start to increase $k_{\rm{F}}$, we find a new type of scaling law related
to the threshold behavior as function of $k_{\rm{F}}$. In figure~\ref{spec}c, we plot
the flow of the critical points, $a_{n}^{-}$, where a three-body bound
state appears from the three-atom continuum at zero energy as $k_{\rm{F}}$ is increased. These are marked with the red dots in figure~\ref{spec}b. 
On the far left of figure~\ref{spec}c in the limit $k_{\rm{F}}\to 0$, we see the usual 
scaling relation. As the thresholds, $a_{n}^{-}$, start to move they approach the two-body 
dimer threshold (black solid line in figure~\ref{spec}c) in a very systematic
fashion.

The solutions shown in figure~\ref{spec}c are clearly self-similar. This is
a hallmark of universal behavior.
We find that a scaling law governs the relation between $a_{n}^{-}$
and $k_{\rm{F}}$. The $k_{\rm{F}}$ values for which 
an Efimov trimer threshold crosses resonance (nearly vertical 'jumps' in 
figure~\ref{spec}c) are given by the universal
relation
\begin{equation}
k_{\rm{F}}^{(n)}=\alpha e^{n\pi/s_0}k^*,
\end{equation}
where $\alpha$ is a non-universal constant that sets the overall scale.
Recall that this relation connects a feature of the background, $k_{\rm{F}}$, 
and a feature of the universal, vacuum three-body system, $s_0$. 
Considering 
that Pauli blocking should influence low-momentum states, while the 
scaling laws are derived from the high-momentum behaviour \cite{braham2006}, 
our results are highly non-trivial. The usual derivation of the 
Efimov scaling relation in vacuum 
will in fact be hindered 
by an obstructing term at low momentum, and the recovery of the scaling 
form is not obvious. 
The scaling 
with $k_{\rm{F}}$ is also found for the points at which the trimer merges with the
atom-dimer continuum as seen in figure~\ref{spec}c.
When considering either the ratios of the trimer energies on resonance 
(where Efimov originally found the infinite tower of universal states), or
for the critical scattering length at which the trimers merge with the 
atom-dimer continuum at non-zero three-body energy 
(the black dots in figure~\ref{spec}b) we again find the 
same behavior when varying $k_{\rm{F}}$. The scaling on resonance 
was obtained previously in Ref.~\cite{macneill2011} but only in the limit of 
very large mass difference. While the results
presented here have assumed equal mass particles, we find the same behavior
for different masses when $s_0$ is changed accordingly. We address the 
advantages of using different mass ratios below.

Above we assume an inert Fermi sea. 
This is expected to be a good approximation on the 
$a<0$ side of the resonance.
When interactions are present one must in general consider the 
effects of exciting particle-hole pairs in the Fermi sea. In principle, 
this can be done in a self-consistent manner through dressed
particle-hole propagators which is numerically very difficult. 
The formalism can, however, be used to take corrections
to the inert sea into account order by order (similar to the 
discussion in Ref.~\cite{combescot2}).
To estimate these effects, we have used a $k_{\rm{F}}$-dependent 
effective scattering length that includes the well-known 
Gorkov-Melik-Barkudarov particle-hole fluctuations  
(see Appendix~\ref{app-higher}). 
We find that while the positions of thresholds will move, the scaling law discussed
above remains the same. 
This suggests that our findings could be
robust even when including particle-hole correlations, and indicates the 
generic nature of Efimov scaling even in the presence of a Fermi sea. However, 
there are some intrinsic differences between the two- and three-body problems
which can for instance be seen in the rather smooth behavior of the two-body
wave function through the BCS-BEC crossover in contrast to the oscillatory
nature of the Efimov three-body wave function. Further studies of the influence
of particle-hole correlations on the Efimov effect are therefore necessary
to test its robustness in this type of many-body environment.
Furthermore, we caution that our estimates of the particle-hole 
effects will become much worse and eventually break down on the 
BEC side of the resonance where strongly bound dimers are present in 
the system. Monte Carlo calculations \cite{monte} find a two-body dimer 
threshold that is larger than results from low-order diagrammatical calculations \cite{zhou}. 
While it has been shown that when including the effects of single
particle-hole pairs, the result is in very good agreement with Monte Carlo
calculations even on the BEC side
\cite{chevy,combescot2,combescot1,mora,punk} (see Appendix~\ref{app-higher}), 
this does not necessarily imply similar good agreement for the more complicated 
Efimov three-body systems and further studies are warranted.
In any case, for the observation of Efimov physics and the results 
discussed here it is not necessary to approach the deep BEC regime.

\begin{figure*}[ht!]
\centering
\includegraphics[scale=0.65,clip=true]{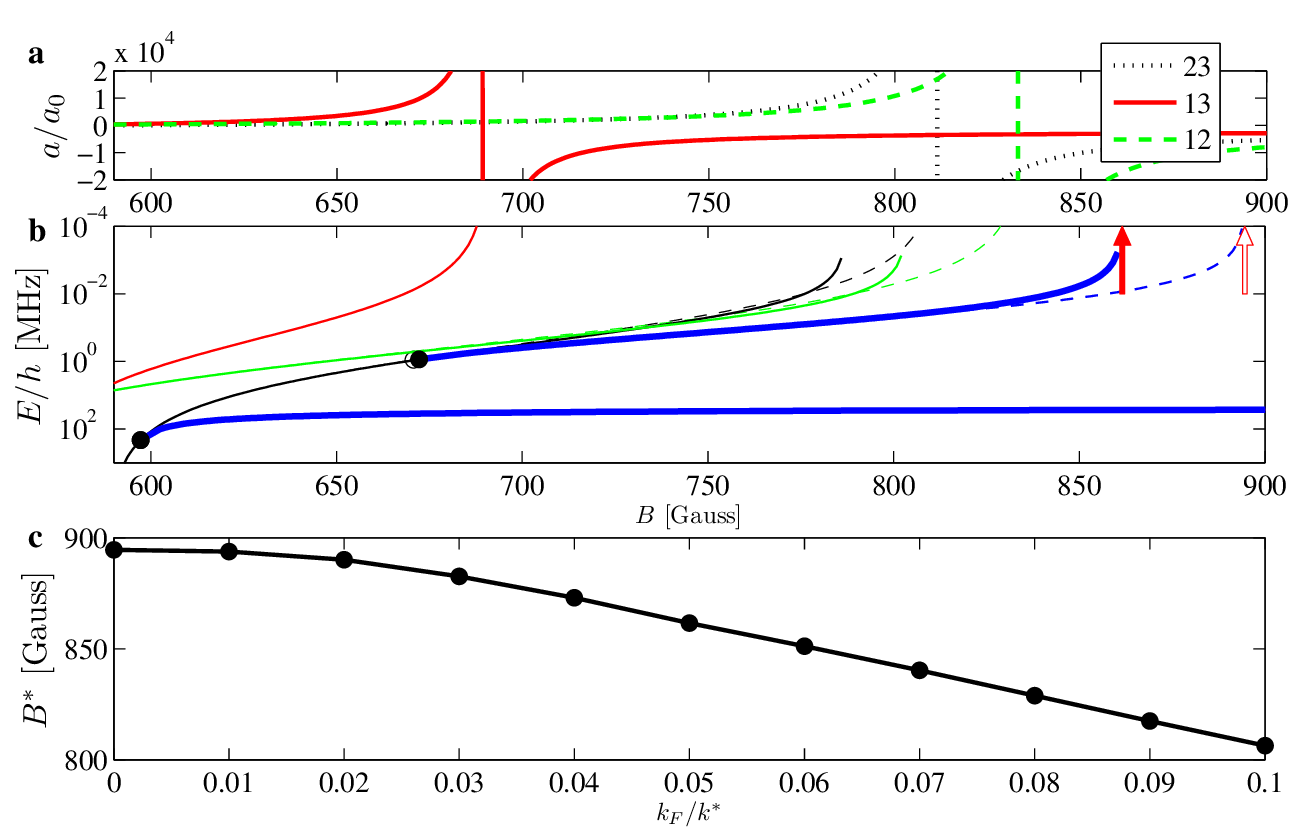}
\caption{Trimer spectrum for a three-component $^{6}$Li system in the presence of a Fermi sea. {\bf a} 
Scattering length and {\bf b} energies as function of magnetic field when state
$|2\rangle$ has a Fermi sea with $k_{\rm{F}}=0.05k^*$ (using labels from Ref.~\cite{lompe2010}). The dashed lines are the results with no
Fermi sea. Black dots indicate the intersections of the trimers with the atom-dimer thresholds. The threshold of the second trimer state in the 
presence of a Fermi sea is 
marked by a red arrow, while a white arrow indicates the position with no Fermi background.
{\bf c} Magnetic field position, $B^*$, of the second Efimov trimer threshold (red arrow in {\bf b})
as $k_{\rm{F}}$ is increased.\label{6li}
}
\end{figure*}

We now turn to the experimental signatures of a many-body background on 
Efimov physics. The signature we will discuss is the loss peak observed
at the threshold for three-body bound state formation from the three-atom
continuum on the $a<0$ side of a Feshbach resonance. Our assumption will 
be that this peak is moved as the Fermi sea modifies the binding energy
of the three-body state and thus the threshold value. A more precise way
would be to include a (small) imaginary part in the three-body parameter
or in the real-space three-body potential to simulate the loss 
\cite{braham2006,loss2013}. This can typically produce both a peak 
position and shape. However, here we focus only on the peak and assume 
that it follows the threshold as in the vacuum case. The question of what 
happens to the shape of the peak in the presence of a Fermi sea will be 
left for future studies.

Presently, the experimentally most relevant system is 
the three-component $^{6}$Li gas where signatures of Efimov states have
been observed by a number of groups in mixtures comprised of the three lowest atomic 
hyperfine states, conventionally labelled $|1\rangle$, $|2\rangle$ 
and $|3\rangle$ \cite{huckans2009,lompe2010,nakajima2011}. The variation of 
the scattering lengths $a_{ij}$ with an applied magnetic field exhibits three 
broad Feshbach resonances as shown in figure~\ref{6li}a.
In figure~\ref{6li}b and c, we present results for a Fermi sea
in hyperfine state $|2\rangle$. We predict a clear displacement of the
magnetic field position at which the second Efimov trimer merges with the continuum, which 
becomes more pronounced as $k_{\rm{F}}$ increases. This state has been observed
as a resonant three-body loss peak at densities $5\times 10^9$ atoms/cm$^{3}$ at a temperature 
of 30 nK \cite{williams2009}
and used to determine the three-body parameter 
$k^*=6.9(2)\times 10^{-3}\,a_{0}^{-1}$ ($a_0=5.29\cdot 10^{-11}$ m).
We use this value of $k^*$ in figure~\ref{6li}, which can be converted to a 
density through $k_{\rm{F}}/k^*=3\cdot 10^{-6} (n/\textrm{cm}^{-3})^{1/3}$.
Notice that the trimer state we consider is on the BCS side of the resonance
and we expect particle-hole correlation effects to be small as discussed above.
We estimate that the density used in Ref.~\cite{huckans2009} to be too low to 
see the effect of the Fermi sea (see figure~\ref{6li}c). 
Therefore the experiment provides a consistent estimation of $k^*$. While larger densities have been 
used \cite{huckans2009,nakajima2011,otten2008}, no results for the 
second trimer at large density have been reported.

In order for the presence of the many-body background to move the three-body loss resonance in 
the $^6$Li system by about 40 Gauss as shown in figure~\ref{6li}b, we estimate 
that densities of $5\cdot 10^{12}$ atoms/cm$^{3}$ are required. In light of
uncertainties in $k^*$ \cite{williams2009,nakajima2010}, 
this estimate could increase by an order of magnitude. A potential problem of 
measuring the loss signal at finite $k_{\rm{F}}$ is that the trimer moves closer
to the Feshbach resonance where the scattering length is very large. This reduces the 
so-called threshold regime where the loss feature is large \cite{incao2004}. 
The Fermi energy can reduce the observability in similar fashion. However, 
more recent results \cite{wang2011} indicate that universal physics is accessible
even at finite collisional two-body energies. We expect this to hold true 
when adding the degenerate Fermi sea and that the movement of the 
loss peak from the second trimer state should be observable.
Changing the energy of the first trimer on the other side of the resonance
where a dimer exists and photo association is possible \cite{lompe2010,nakajima2011}
requires unrealistically high densities. We note that the differences in the $^{6}$Li 
scattering lengths can be reduced dramatically by applying additional optical
fields \cite{ohara2011}, bringing the real system closer to the generic Efimov
system considered above.

The $^{6}$Li system is in a sense not an optimal choice for observing the 
flow of the Efimov spectrum as equal mass systems has a large scaling factor, $e^{2\pi/s_0}$,
so the trimer states are far apart in energy. On the other hand, if the 
mass ratio(s) are large, the spectrum is much denser \cite{jen03}. Additionally,
heteronuclear Feshbach resonances tend to have smaller
background scattering lengths \cite{chin2010} which could be beneficial 
in order to not saturate the unitary limit and wash out the loss peak(s). Since we find that the 
movement of trimer energies with $k_{\rm{F}}$ and the scaling laws are generic
in the universal limit, we expect that hetero-atomic mixtures are a 
much better options for observing these background effects when the 
fermionic component(s) has large density. We note that a very recent
experiment at MIT \cite{wu2011} using mixtures of Bose and Fermi species fit
very well with the requirements to make our predictions more
readily observable. While we have focused on trimer energies, the 
STM equations are perfectly suited
for calculating the loss rates \cite{braham2006,braaten2009} and our work demonstrates
how one can include effects of a degenerate background in these quantities.

Here we have only considered the case of a Fermi sea in a single component
of the trimer system. As the three-body Efimov physics depends 
mostly on two-body subsystem thresholds, we expect to find qualitatively similar results
with additional Fermi seas (see discussion in the supplementary material).
We do expect that with two Fermi seas one could have
Cooper pairs for any $a$, and in turn the dimer threshold would go to $a\to 0^-$. This 
extends the atom-dimer continuum and one can imagine that the trimer thresholds
also move toward $a\to 0^-$. However, the binding energy
of Cooper pairs is exponentially small in that limit and there could still
be a critical value $a^-<0$ for forming trimers. An additional question
is the effect of superfluidity at low temperatures. This will 
presumably also change the spectrum, but we doubt that the scaling relation 
with $k_{\rm{F}}$ is modified. We note that many-body effects on trimers
should be present in bosonic samples as well \cite{bosons2013} and will in
some cases cause similar bound state energy spectral flow with background 
parameters (condensate density) \cite{flow2014} as seen in the present case with fermions. Recent experiments
on $^7$Li with thermal and condensed samples find apparent discrepancies \cite{pollack2009,gross2010}, which
could be attributed to effects of the finite range of the inter-atomic interactions \cite{dyke2013}.
It would be interesting to investigate whether the many-body background could also
play a role. The approach
pursued here should be applicable to bosonic systems as well, an obvious
direction for future work. Lastly, the influence of a background environment
on the recently measured four-body states \cite{ferlaino2009,pollack2009} 
is an interesting open question.

\ack
We are grateful to A.S. Jensen, G.M. Bruun, D.V. Fedorov and T. Lompe for extensive discussions, and to 
I. Zapata and K. M{\o}lmer for comments on the manuscript. We also thank
E. Demler, B. Wunsch, F. Zhou and M. Ueda for useful comments and feedback on the results.

\appendix

\section{\bf Dressed Dimer Propagator.}
The dressed dimer can be obtained from the Bethe-Selpeter equation in the ladder approximation \cite{fetter}. 
Here we consider only a zero-range interaction, which amounts to a constant in momentum space. This 
leads to a dimer propagator (or two-body $T$-matrix) between atoms $i\neq j$ of the form
\begin{equation}
D_{ij}({\bf k},\omega) = T_{ij}({\bf k},\omega) = \frac{g_{ij}}{1-g_{ij}\Pi_{ij}({\bf k},\omega)},
\end{equation}
where $\Pi_{ij}({\bf k},\omega)$ is the pair bubble and $g_{ij}$ is the coupling constant of the 
zero-range potential. Introducing a Fermi sea to be in component $j$ with Fermi wave number $k_{\rm{F}}$, 
the pair bubble at zero temperature 
becomes
\begin{eqnarray}\label{bubble}
\Pi_{ij}({\bf k},\omega) 
&=&  \int \frac{d^3q}{(2\pi)^3} \, \frac{\Theta(q-k_{\rm{F}})}{\hbar\omega-\frac{\hbar^2q^2}{2m_j}-\frac{\hbar^2({\bf k}+{\bf q})^2}{2m_i}+\mu_j+i\epsilon},
\end{eqnarray}
where $m_i$ and $m_j$ are the particle masses, $\mu_j=\hbar^2k^2_{\rm{F}}/2m_j$ the chemical potential of component $j$ and
$\epsilon$ is an infinitesimal positive number. At zero temperature, we see that 
the dimer propagator is modified by a restriction of Hilbert space due to Pauli blocking in one of the components 
as the Fermi sea is introduced. The usual linear divergence at large momentum with the momentum cut-off, $\Lambda$, 
persists when modifying the Hilbert space at low momentum through $k_{\rm{F}}$. This is renormalized through the 
two-body scattering length $a_{ij}$ which can be related to the vacuum ($k_{\rm{F}}=0$) $T$-matrix, $T^0_{ij}$, by
\begin{equation}
a_{ij} = \frac{m_{ij}}{2\pi\hbar^2} T^0_{ij}(0,0)
= \frac{\frac{m_{ij}}{2\pi\hbar^2}}{g^{-1}_{ij}+\frac{m_{ij}\Lambda}{\pi^2\hbar^2}},
\end{equation} 
where $m_{ij}=m_i m_j/(m_i+m_j)$ is the reduced mass of the dimer.  The renormalized dimer propagator is 
\begin{eqnarray}
&D_{ij}({\bf k},\omega)=
\frac{2\pi\hbar^2}{m_{ij}} \frac{1}{a_{ij}^{-1}-\frac{\sqrt{2m_{ij}}}{\hbar}\sqrt{-\hbar\omega+\frac{\hbar^2k^2}{2M_{ij}}-\mu_j-i\epsilon}+R({\bf k},\omega)},&\label{dimer}
\end{eqnarray}
where the contribution of the Fermi sea in component $j$ is seen in the chemical potential, $\mu_j$, and the 
many-body correction term 
\begin{eqnarray}
&R({\bf k},\omega) =
\frac{m_i k}{2\pi m_{ij}}\biggl\lbrace
2\frac{m_{ij}}{m_i} x_{\rm{F}}-\frac{x_{\rm{F}}^2}{2} \ln \left[ \frac{(x_{\rm{F}}-x_+)(x_{\rm{F}}-x_-)}{(x_{\rm{F}}+x_+)(x_{\rm{F}}+x_-)} \right]&
\nonumber \\
& -\frac{x_+^2}{2} \ln \left( \frac{x_{\rm{F}}+x_+}{-x_{\rm{F}}+x_+} \right) -\frac{x_-^2}{2} \ln \left( \frac{x_{\rm{F}}+x_-}{-x_{\rm{F}}+x_-} \right) \biggr\rbrace,&
\label{R}
\end{eqnarray}
with $x_{\rm{F}}=\frac{k_{\rm{F}}}{k}$ and 
\begin{equation}
x_\pm=\frac{m_{ij}}{m_i}\pm\sqrt{\left(\frac{m_{ij}}{m_i}\right)^2+\frac{2m_{ij}(\hbar\omega+\mu_j)}{\hbar^2k^2}-\frac{m_{ij}}{m_{ik}}+i\epsilon}.
\end{equation}

The bound state pole of equation (\ref{dimer}) is used to determine the boundary of the atom-dimer continuum shown in figure~1b and 1c of the main text. 
We note that in the case of equal masses, the lowest energy state of the dimer is always at 
zero center-of-mass momentum. In fact, this is true whenever $m_i\geq m_j$ with $j$ having the Fermi sea. However, this is 
not a necessary assumption in our setup which works just as well for $m_i<m_j$ where non-zero center-of-mass dimers may be
lower in energy \cite{schuck,combescot2,songs}.
The critical scattering length, $a_{ij}^{c}$ at which the dimer pole appears at zero energy, $E=0$, in the presence of a Fermi sea has the 
analytical expression 
\begin{equation}
\frac{1}{a_{ij}^{c}} = \frac{2k_{\rm{F}}}{\pi} \left( 1-\frac{1}{2}\sqrt{\frac{m_{ij}}{m_j}}\ln \left| \frac{1+\sqrt{m_j/m_{ij}}}{1-\sqrt{m_j/m_{ij}}} \right| \right).
\label{continua_intersect}
\end{equation}
In the equal mass case we have $\frac{1}{a_{ij}^{c}}=0.24k_F$ which marks the edge of the atom-dimer continuum in figure~1c of the main text.

\subsection{Higher-order Corrections}\label{app-higher}
Here we discuss an estimate of the effect of particle-hole 
pairs in the two-body dimer, i.e. of a system of two-component
fermions. As the two-body propagator 
enters the calculation of the three-component system it should 
yield an insight into how this affects the three-body states.

The dimer propagator in (\ref{dimer}) is obtained from the ladder approximation to lowest order. At next order, we need to include 
the effect of particle-hole excitations in the Fermi sea in analogy to what has been discussed in fermionic superfluids by Gor'kov and Melik-Barkhudarov \cite{gmb}. 
As shown in Ref.~\cite{zhou}, this can be achieved by using an effective scattering
length, $\tilde a_{ij}$, given by 
\begin{equation}
\tilde a_{ij}^{-1}=a_{ij}^{-1}-Fk_{\rm{F}},
\end{equation} 
where $F$ contains the lowest order vertex correction
and depends on the mass ratio $m_i/m_j$. For $m_i=m_j$, $F\sim 0.2$. To check the robustness of our finding we have calculated 
trimer energies with $a_{ij}$ replaced by $\tilde a_{ij}$, and we have also varied the value of $F$. 
While this modifies the absolute values of the binding energies, it does not alter the scaling relation in $k_{\rm{F}}$. 
We therefore believe that the scaling is unaltered by vertex corrections.

Diagrammatic Monte Carlo studies of the strongly imbalanced two-component Fermi system (the so-called polaron problem) \cite{monte} does, however,
show that the critical scattering length is $(k_{\rm{F}}a_{ij}^{c})^{-1}=0.88$, which is higher than both the lowest order ladder 
approximation ($(k_{\rm{F}}a_{ij}^{c})^{-1}=0.24$) and the result including particle-hole correlations ($(k_{\rm{F}}a_{ij}^{c})^{-1}=0.34$). 
As noted in Ref.~\cite{zhou}, this implies that the vertex corrections are qualitatively correct but underestimate the effect of 
higher-order corrections. In the polaron problem it has been shown that inclusion of one particle-hole pair correlations 
on top of the inert Fermi sea gives results that are similar to the Monte Carlo findings \cite{combescot1,combescot2,mora,punk}.
In a $T$-matrix approach such as the one used here, these corrections can be taken systematically into account through self-energy
terms.

We note that the dimer propagator in (\ref{dimer}) is not accurate in the deep BEC limit ($a\to 0^+$) where
it can be shown to yield the Born approximation for the atom-dimer scattering length, $a_3/a_{ij}=8/3$, which overestimates the 
exact result of $a_3/a_{ij}=1.18$ originally obtained in Ref.~\cite{STM} and subsequently reproduced and discussed by several authors, see for instance Ref.~\cite{combescot2}. We believe that all these corrections should be only quantitative
a more detailed analysis that takes the self-energy term in the single-particle propagators and higher-order terms in the 
dimer propagator is necessary to confirm this. 
Quantitatively, we expect the present treatment to work well for $a_{ij}<0$ and
around the resonance, $1/a_{ij}\to 0$. However, observing the Efimov scaling in $k_{\rm{F}}$ does not require calculations in 
the deep BEC limit as demonstrated in figure~1 of the main text and our findings should be recovered in more involved treatments.

\subsection{Additional Fermi seas}
The considerations above pertain to the case where one of the particles that make up the three-body bound state has a background
Fermi sea. A natural extension is to consider two or three Fermi seas. The STM equations that we use can perfectly well 
accommodate such a system when properly modifying the dimer propagator. For the case of two Fermi seas, the pair bubble of (\ref{bubble})
must be replaced by
\begin{eqnarray}
&\Pi_{ij}({\bf k},\omega)= 
\int \frac{d^3q}{(2\pi)^3} \, \frac{1-\Theta(k_{\rm{F}}-q)-\Theta(k_{\rm{F}}-|{\bf q}+{\bf k}|) }{\hbar\omega-\frac{\hbar^2q^2}{2m_j}-\frac{\hbar^2({\bf k}+{\bf q})^2}{2m_i}+\mu_i+\mu_j+i\epsilon},&
\end{eqnarray}
where we assume two inert seas and thus propagation of a pair of particles above the Fermi seas only. We are 
also assuming that the two seas have the same size, $k_{\rm{F}}$, but this is not essential and can be easily 
generalized to imbalanced systems. The pair propagator above must be computed numerically in general. This
form of the propagator appears for instance in the original work of Galitskii on the Fermi gas with strong
short-range interactions \cite{fetter,galitskii}.

The goal of the current section is, however, to argue that the Efimov scaling in $k_{\rm{F}}$ discussed in the 
main text is maintained in the case of two Fermi seas. This can be done by considering the structure of the 
pair bubble in the limit where the center-of-mass momentum, ${\bf k}$, vanishes. Here we obtain
\begin{eqnarray}
&D_{ij}({\bf 0},\omega) =
\frac{2\pi\hbar^2}{m_{ij}} \frac{1}{a_{ij}^{-1}-\frac{\sqrt{2m_{ij}}}{\hbar}\sqrt{-\hbar\omega-\mu_i-\mu_j-i\epsilon}+2R({\bf 0},\omega)},&
\end{eqnarray}
in the limit ${\bf k}\to 0$. In the expression for $R({\bf 0},\omega)$ one must also make the substitution
$\mu_j\to \mu_j+\mu_i$. This holds in the equal 
mass case and for an unpolarized system with Fermi seas in both components. In particular, this has the same
structure as the case of a single Fermi sea discussed above. We have checked numerically that multiplication of 
$R({\bf 0},\omega)$ by two does not alter the scaling relations discussed in the main text. In the limit
of small $k_{\rm{F}}$, $R({\bf 0},\omega)$ is proportional to $k_{\rm{F}}$. This shows that the changes
required from addition of a second Fermi sea are similar to those coming from Gor'kov-Melik-Barkhudarov corrections
above. While threshold will change, the overall scaling properties remain the same. We note that 
the shift of the critical value for a two-body bound state in the presence of a 
Fermi sea to the $a_{ij}>0$ side of the Feshbach resonance continues to hold true with Fermi seas
in both components \cite{combescot3}.

The argument given was for the case of vanishing center-of-mass momentum of the pairs in the dimer propagator.
It is known from the Cooper pair problem that the strongest bound pairs have zero center-of-mass momentum \cite{fetter}.
Thus for the equal mass three-body states we expect this dimer pole to dominate, validating the arguments
given above. Extension to three Fermi seas is straightforward and since the propagators are the same
the same arguments can be used to infer that the scaling should persist there as well.
If one considers the interesting case of unequal mass systems this may no longer be the case
and the strongest bound two-body state may acquire non-zero center-of-mass momentum. All this can be 
studied within our formalism and we are actively pursuing this direction at the moment.

\section{\bf Momentum-space Three-body Equations.}
The momentum-space three-body equations are also known as the Skorniakov-ter-Martirosian equations (STM) \cite{STM} and 
where originally developed for problems in nuclear physics. 
By a ladder summation over all pairwise interactions the integral equations for the three-body $T$-matrix can be casts as the scattering of a dimer with an unbound atom. In a three component mixture there are thus three entrance channels and three exit channels for the free atom giving a $3\times 3$ $T$-matrix $T_{ij}$ \cite{braaten2010}.	
The equations for the three-body scattering amplitudes, $T_{ij}(p,q)$, with a single Fermi sea in component $3$ after 
angular-averaging to get the zero angular momentum part are ($i=1,2,3$)
\begin{eqnarray}
&T_{i1}(p,q) = \sum_{k=1}^3 (1-\delta_{i1})(1-\delta_{ik})(1-\delta_{1k})K_{i1}^k(p,q)& \nonumber \\
&+ \frac{1}{4\pi^2} \int_{-1}^1 d(\cos\theta) \int_0^\Lambda dQ \,Q^2 \times  &\nonumber\\
&\frac{2m_{23}\Theta(Q^2+q^2+2Qq\cos\theta-k_{\rm{F}}^2) D_{13}\left({\bf Q},E-\frac{\hbar^2Q^2}{2m_2}\right)T_{i2}(p,Q)}{2m_{23}E-\hbar^2Q^2-\frac{m_{23}}{m_{13}}\hbar^2q^2-2\frac{m_{23}}{m_3}\hbar^2Qq\cos\theta+i0^+}&
\nonumber \\
&-\frac{1}{2\pi^2}   \int_{k_{\rm{F}}}^\Lambda dQ \,  Q^2 K_{31}^2(Q,q)D^0_{12}\left({\bf Q},E-\frac{\hbar^2Q^2}{2m_3}\right)T_{i3}(p,Q),&\\
&T_{i2}(p,q) = \sum_{k=1}^3 (1-\delta_{i2})(1-\delta_{ik})(1-\delta_{2k})K_{i2}^k(p,q)&\nonumber \\
&+ \frac{1}{4\pi^2} \int_{-1}^1 d(\cos\theta) \int_0^\Lambda dQ \,Q^2  \times&\nonumber\\
& \frac{2m_{13}\Theta(Q^2+q^2+2Qq\cos\theta-k_{\rm{F}}^2) D_{23}\left({\bf Q},E-\frac{\hbar^2Q^2}{2m_1}\right)T_{i1}(p,Q)}{2m_{13}E-\hbar^2Q^2-\frac{m_{13}}{m_{23}}\hbar^2q^2-2\frac{m_{13}}{m_3}\hbar^2Qq\cos\theta+i0^+}& 
\nonumber \\
&-\frac{1}{2\pi^2}   \int_{k_{\rm{F}}}^\Lambda dQ \,  Q^2 K_{32}^1(Q,q)D^0_{12}\left({\bf Q},E-\frac{\hbar^2Q^2}{2m_3}\right)T_{i3}(p,Q),&
 \\
&T_{i3}(p,q) = \sum_{k=1}^3 (1-\delta_{i3})(1-\delta_{ik})(1-\delta_{3k})K_{i3}^k(p,q)&
\nonumber \\
&-\frac{1}{2\pi^2}   \int_{0}^\Lambda dQ \,  Q^2 K_{13}^2(Q,q)D_{23}\left({\bf Q},E-\frac{\hbar^2Q^2}{2m_1}\right)T_{i1}(p,Q)&
\nonumber \\
&-\frac{1}{2\pi^2}   \int_{0}^\Lambda dQ \,  Q^2 K_{23}^1(Q,q)D_{13}\left({\bf Q},E-\frac{\hbar^2Q^2}{2m_2}\right)T_{i2}(p,Q),&
\end{eqnarray}
where the Heaviside step function $\Theta(x)$ is 1 for $x>0$ and vanishes for $x<0$ and $E$ is the three-body energy. 
The superscript '0' on the dimer propagator $D^0_{12}$ is a reminder that it does not contain a Fermi sea and is the vacuum expression 
which can be recovered from (\ref{dimer}) by setting $\mu_j=0$ and $R({\bf k},\omega)=0$. The kernel, $K_{ij}^{k}(p,q)$, is defined by
\begin{equation}
K_{ij}^k(p,q) = \frac{m_k}{2\hbar^2 pq} \ln \left[ \frac{p^2+\frac{m_{ik}}{m_{jk}}q^2+2\frac{m_{ik}}{m_k}pq-\frac{2m_{ik}E}{\hbar^2}}{p^2+\frac{m_{ik}}{m_{jk}}q^2-2\frac{m_{ik}}{m_k}pq-\frac{2m_{ik}E}{\hbar^2}} \right].
\end{equation} 

Since we are only interested in the spectrum of bound states in the present work, we neglect the homogeneous terms and use a standard pole expansion technique \cite{taylor}
by which we can write
\begin{equation}
T_{ij}(p,q)=\frac{B_i(p)B_j(q)}{E-E_T},
\end{equation}
where $E_T$ is the trimer energy that we want to determine, and $B_i(p)$ are the residues at the poles, which represent the three-component bound state wave function. This yields the bound state equations
\begin{eqnarray}
&B_{1}(p) =  \frac{1}{4\pi^2} \int_{-1}^1 d(\cos\theta) \int_0^\Lambda dq \,q^2   \times&\nonumber\\
&\frac{2m_{23}\Theta(q^2+p^2+2qp\cos\theta-k_{\rm{F}}^2) D_{13}\left({\bf q},E-\frac{\hbar^2q^2}{2m_2}\right)B_{2}(q)}{2m_{23}E-\hbar^2q^2-\frac{m_{23}}{m_{13}}\hbar^2p^2-2\frac{m_{23}}{m_3}\hbar^2qp\cos\theta+i0^+}&
\nonumber \\
&-\frac{1}{2\pi^2}   \int_{k_{\rm{F}}}^\Lambda dq \,  q^2 K_{31}^2(q,p)D^0_{12}\left({\bf q},E-\frac{\hbar^2q^2}{2m_3}\right)B_{3}(q),&
\\
&B_{2}(p) = \frac{1}{4\pi^2} \int_{-1}^1 d(\cos\theta) \int_0^\Lambda dq \,q^2   \times&\nonumber\\
&\frac{2m_{13}\Theta(q^2+p^2+2qp\cos\theta-k_{\rm{F}}^2) D_{23}\left({\bf q},E-\frac{\hbar^2q^2}{2m_1}\right)B_{1}(q)}{2m_{13}E-\hbar^2q^2-\frac{m_{13}}{m_{23}}\hbar^2p^2-2\frac{m_{13}}{m_3}\hbar^2qp\cos\theta+i0^+} &
\nonumber \\
&-\frac{1}{2\pi^2}   \int_{k_{\rm{F}}}^\Lambda dq \,  q^2 K_{32}^1(q,p)D^0_{12}\left({\bf q},E-\frac{\hbar^2q^2}{2m_3}\right)B_{3}(q),&
\\
&B_{3}(p) = -\frac{1}{2\pi^2}   \int_{0}^\Lambda dq \,  q^2 \times &\nonumber\\
&K_{13}^2(q,p)D_{23}\left({\bf q},E-\frac{\hbar^2q^2}{2m_1}\right)B_{1}(q)&
\nonumber \\
&-\frac{1}{2\pi^2}   \int_{0}^\Lambda dq \,  q^2 K_{23}^1(q,p)D_{13}\left({\bf q},E-\frac{\hbar^2q^2}{2m_2}\right)B_{2}(q).&
\end{eqnarray}

The original STM equations are divergent at large momenta and a cut-off, $\Lambda$, is necessary on the integrals. This is caused by the well-known Thomas collapse \cite{thomas1935}.
The equations therefore need to be regularized which can be done in a number of different ways \cite{braham2006}. Here we 
use the elegant subtraction technique of Pricoupenko \cite{pri}, which draws inspiration from the older work of 
Danilov \cite{danilov} on how to impose conditions on the asymptotic (high-momentum) behaviour. 
After regularization, the equations become
($i\neq j \neq k$)
\begin{eqnarray}
&B_{i}(p) = -\frac{1}{2\pi^2} 
\int_0^\Lambda dq \, q^2 \times&\nonumber\\
&\left[ \Delta {\mathcal{K}}_{ki}^{j}(q,p)D_{ij}\left({\bf q},E-\frac{\hbar^2q^2}{2m_k} \right) B_{k}(q) \right.&\nonumber \\
&\left. + \Delta {\mathcal{K}}_{ji}^{k}(q,p)D_{ik}\left({\bf q},E-\frac{\hbar^2q^2}{2m_j} \right) B_{j}(q)\right],&
\end{eqnarray}
where $\Delta {\mathcal{K}}_{ij}^k(q,p)={\mathcal{K}}_{ij}^k(q,p)-{\mathcal{K}}_{ij}^k(q,k_{\rm{reg}})$, and we have defined
\begin{eqnarray}
&{\mathcal{K}}_{3j}^k(q,p) = K_{3j}^k(q,p)\Theta(q-k_{\rm{F}}),& \\
&{\mathcal{K}}_{i3}^k(q,p) = K_{i3}^k(q,p),& \\
&{\mathcal{K}}_{ij}^3(q,p) = -\frac{1}{2} \int_{-1}^1 d(\cos\theta)\times &\nonumber\\ &\frac{2m_{i3}\Theta(q^2+p^2+2qp\cos\theta-k_{\rm{F}}^2)}{2m_{i3}E-\hbar^2q^2-\frac{m_{i3}}{m_{j3}}\hbar^2p^2-2\frac{m_{i3}}{m_3}\hbar^2qp\cos\theta+i0^+}.&
\end{eqnarray}
The parameter in the regularizing part of the kernel can be expressed as \cite{pri} 
\begin{equation}
\frac{k_{\rm{reg}}}{k^*}=\frac{1}{\sqrt{3}}e^{\pi(n+1)/s_0}, 
\end{equation}
where $k^*$ is the three-body parameter which the basic unit in the three-body problem and which
cannot be determined within the universal theory containing only scattering length parameters \cite{braham2006}. Here
$s_0$ is the scale factor and $n$ is a normalization factor which fixes the minimum energy of trimer states \cite{pri}. In 
all our numerical calculations we use $n=0$. 
The regularization scheme is derived in the large momentum
limit where the equations become scale invariant. The inclusion of a Fermi sea does not alter the asymptotic 
behavior and the same regularization can therefore be used in this case also.

\end{document}